\documentstyle[twoside,fleqn,espcrc2,epsfig]{article}

\def\spose#1{\hbox to 0pt{#1\hss}}
\def\lsim{\mathrel{\spose{\lower 3pt\hbox{$\mathchar"218$}}
 \raise 2.0pt\hbox{$\mathchar"13C$}}}
\def\gsim{\mathrel{\spose{\lower 3pt\hbox{$\mathchar"218$}}
 \raise 2.0pt\hbox{$\mathchar"13E$}}}

\begin{document}

\begin{titlepage}

\begin{flushright}
DESY-98-106\\
hep-ph/9808329
\end{flushright}

\vspace{1.5cm}


\boldmath
\begin{center}
\Large\bf Charmless Nonleptonic Two-Body $B$
 Decays in the Factorization 
Approach
\end{center}
\unboldmath

\vspace{1.2cm}

\begin{center}
 Cai-Dian L\"u\footnote{Alexander 
von Humboldt research  fellow.}

{\sl  II Institut f\"ur Theoretische Physik, Universit\"at Hamburg,
D - 22761 Hamburg, Germany
}
\end{center}

\vspace{1.3cm}

\begin{center}
{\bf Abstract}\\[0.3cm]
\parbox{11cm}{
We calculate the two-body nonleptonic B decays using the factorization
method. The recent 
measured decays by CLEO Collaboration can be explained in the factorization
approach. We propose a number of ratios of branching ratios
 to determine the effective coefficients and the form factors.

}

\end{center}

\vspace{1cm}

\begin{center}
{\sl Invited talk given at the\\
6th International Euroconference QCD 98\\
Montpellier, France, 2--8 July 1998\\
To appear in the Proceedings}
\end{center}

\vspace{1.5cm}



\end{titlepage}

\thispagestyle{empty}
\vbox{}
\newpage

\setcounter{page}{1}



\title{ Charmless nonleptonic two-body $B$
 decays in the factorization  approach
}

\author{ 
C.-D. L\"u\\
II Institut f\"ur Theoretische Physik, Universit\"at Hamburg,
 22761 Hamburg, Germany\thanks{Alexander von Humboldt research  
fellow.}}

\begin{abstract}
We calculate the two-body nonleptonic B decays using the factorization
method. The recent 
measured decays by CLEO Collaboration can be explained in the factorization
approach. We propose a number of ratios of branching ratios
to determine the effective 
coefficients and the form factors from experiments.
\end{abstract}

\maketitle

\section{INTRODUCTION}

In this report I represent works done in collaboration with A. Ali and 
G. Kramer \cite{akl1}.

Recently the CLEO Collaboration has measured a number of  
charmless nonleptonic  B decays \cite{cleo}. This has aroused 
 great interest of studying these decays \cite{ag,deshpande,others}. 
The charmless nonleptonic  B decays provide the opportunity to 
 study  CP violation and a way of
measuring of CKM matrix elements.
Since they are rare decays, they are also useful for 
tests of the Standard Model and giving
signals of possible new physics.

The calculation of nonleptonic decays involves the short-distance Lagrangian 
and the calculation of hadronic matrix elements which are model dependent.
 The short-distance QCD corrected Lagrangian is  calculated
 to { next-to-leading order}. 
The popular method to calculate the  hadronic matrix elements is 
using the factorization method where the matrix element is expressed as
a product of two factors
 $\langle h_1h_2 | {\cal H}_{eff}|B\rangle =\langle h_1 | J_1|B \rangle 
\langle h_2 | J_2|0 \rangle $. 

This model works well for the tree level hadronic B and D decays using 
phenomenological values of $a_1$ and $a_2$ from experiments \cite{bsw,ns}. 
Does it also work for decays involving light mesons? 
In charmless decays, penguins play an important role. 
What is the effective number of color $N_c$ in this category of decays?
Here we use the factorization method to
analyze the charmless nonleptonic decays of 
76 channels and their charge conjugated decays.
A number of ratios is proposed to test the factorization approach and 
measure the parameters $a_i$ and CKM matrix elements.

\section{ THEORETICAL FRAMEWORK}

The effective Hamiltonian for the charmless  nonleptonic B decays is 
\begin{eqnarray}
\label{heff}
{\cal H}_{eff}
= \frac{G_{F}}{\sqrt{2}} \, \left[ 
V_{q'b} V_{q'q}^* \,
\left(\sum_{i=1}^{10}
C_{i} \, O_i + C_g O_g \right) \right]  ,
\end{eqnarray}
where
$q=d,s$
and $V_{q'q}$ denotes the CKM factors.
The operators $O_1,O_2$ are current operators.
The operators $O_3,\ldots,O_6$ are QCD
penguin operators.
$O_7,\ldots,O_{10}$ arise from 
electroweak penguin diagrams,
which are suppressed by $\alpha/\alpha_s$.
Only $ O_9$ has a sizable value whose 
 major contribution arises from the $Z$ penguin.
In practice, we work
at $ \mu=2.5~GeV $, in the naive dimensional regularization 
scheme.

For example, let us consider B meson decays to two
 pseudoscalar mesons.
In the factorization approach, using the effective Hamiltonian, 
 we  write the required matrix element in its factorized form
{\begin{eqnarray}
  \label{7}
&&\langle P_1 P_2 | {\cal H}_{eff} |  B \rangle \nonumber\\
&&=  i
\frac{G_F}{\sqrt{2}} V_{qb}V_{q q'}^*{ a_i}
f_{P_2} (m_B^2-m_1^2) F_0^{B\to P_1} (m_2^2) .
\end{eqnarray}}
The dynamical details are coded in the  quantities $a_i$, 
which we define as
$a_i \equiv C_i^{eff} + C_j^{eff}/N_c$,
where $\{i,j\}$ is any of the pairs $\{1,2\}$, $\{3,4\}$, $\{5,6\}$,
$\{7,8\}$ or $\{9,10\}$.
In practice,
$N_c$ is treated as a phenomenological parameter to include the 
non-factorized color-octet contributions.

The QCD  and electroweak penguins 
are also present in the charmless decays of B mesons. However,
if the amplitude is 
 still dominated by the tree amplitude, the 
BSW-classification  can be applied as before \cite{bsw}. 
 Class-I decays are color favored, whose matrix elements are 
proportional to $a_1$.
 Class-II decays are color suppressed, whose 
matrix elements are proportional to
 $a_2$.
  Class-III decays are  proportional to
  $a_1 + r a_2$.

For the penguin-dominated decays, we introduce two new  
classes:  Class-IV decays   involve 
one or more of the dominant penguin coefficients
$a_4$, $a_6$ and $a_{9}$.
 Class-V decays are decays with strong $N_c$-dependent  
coefficients $a_3, a_5,a_7$ and $a_{10}$.

Class-I and Class-IV decays have relatively large branching ratios 
of the order of $10^{-5}$ and
stable against variation of 
$N_c$.
 Class-III decays are  mostly stable, except some 
 $B \to PV$ decays.
Class-II and Class-V decays are rather unstable against variation of 
$N_c$. 
 Many of them  may 
receive significant contribution from the annihilation diagrams and/or
soft final state interactions.

\section{NUMERICAL CALCULATIONS}

The CKM matrix is
expressed in terms of the Wolfenstein parameters.
 We take 
$A= 0.81$, $\lambda=\sin \theta_C=0.2205$,
$\rho = 0.12$, $ \eta = 0.34$. 
This choice of $\rho,\eta$ corresponds to 
CKM triangles:
$\alpha =88.3^\circ$, $ \beta = 21.1^\circ$, $\gamma = 70.6^\circ$.
At scale $\mu=2.5 $GeV, we use current masses 
${m_b}=4.88$ GeV, $m_c=1.5$ GeV,
 ${m_s} = 122$ MeV,  
${m_d} = 7.6$  MeV and  ${m_u} = 4.2 $ MeV.
The form factors are taken from ref.~\cite{bsw,akl1} indicated as BSW model 
and Lattice QCD/QCD sum rule.
The decay constants are shown in ref.~\cite{akl1}.

\begin{small}
\begin{table}
\begin{center}
\label{table8}
\caption{ $B\to h_1 h_2$ Branching Ratios (in units of $10^{-6}$). } 
\label{pp1}
\begin{tabular} {lcccc}
\hline
Channel &  $N_c=2$ &  $3$ & $\infty$ & Exp. \\
\hline
$B^0 \to \pi^+ \pi^-$  & $9.0 $ &$10.0 $ 
 & $12  $ & $ < 8.4$ \\
$ B^+ \to \pi^+ \pi^0$  & $6.8 $ &$5.4 $
 &$3.0 $ & $ < 16$\\
$ B^+ \to K^+ \pi^0$   & $9.4 $  & $10 $  &
  $12 $  &$ 15\pm 4\pm 3 $ \\
$B^0 \to K^+ \pi^-$   & $14 $ &$15 $ &
$18 $ &$14\pm 3 \pm 2$\\
$ B^+ \to  K^+ \eta^\prime$   & $21 $  & $25 $  &
  $35 $  &$ 74^{+8}_{-13}\pm 9 $ \\
$B^0 \to K^0 \eta'$   &  $20 $  & $25 $  &
  $35 $  &$ 59^{+18}_{-16}\pm 9 $\\
$ B^+ \to \pi^+ K^0$   & $14 $ & $16 $ 
& $22 $ & $14 \pm 5 \pm 2 $\\
$B^0 \to \rho^+  \pi^- $  & $21$ &  $23 $
 & $28$ & $ <88$
\\
$ B^+ \to \rho^+ \pi^0$    &$14 $ &$13 $
 &$11  $ & $<77$ \\
$ B^+ \to  K^{*0} \pi^+ $   & $5.6 $& $6.9 $
& $10  $ &$<39$\\
$ B^+  \to \omega K^+ $   &   $3.2 $ &    $0.25 $
  &    $11 $  &$ 15^{+7}_{-6}\pm 2$\\
$B^0 \to \rho^+  \rho^-$   &  $18 $&  $20 $
 & $24 $ & $ <2200$\\
\hline
\end{tabular}
\end{center}
\end{table}
\end{small}

We select a few decay  branching ratios  shown in Table~\ref{table8}.
For the whole list of the 76 channels, see ref.~\cite{akl1}.
In this table, it is easy to see that
 {$B^0 \to K^+ \pi^-$, $B^+ \to K^+ \eta^\prime$,
$B^0 \to K^0 \eta^\prime$, $B^+ \to \pi^+ K^0$}, measured by CLEO,
are well explained by factorization.
The experiment of $B^+ \to K^+ \eta^\prime$ favors a small value of 
$\xi=1/N_c$.

The CLEO experiments measured a combined channel
 ${\cal B}(B^+ \to \pi^0 h^+)=(1.6 ^{+0.6}_{-0.5} 
\pm 0.4) \times 10^{-5}$, ($h^+=\pi^+,K^+$). 
 The measurements agree with experiments for all 
possible values of $\xi$.
The CLEO Collaboration also measured 
  ${\cal B}(B^+ \to \omega h^+)= (2.5 
^{+0.8}_{-0.7} \pm 0.3) \times 10^{-5}$.
This suggests
 values of $\xi \simeq 0$ and $\xi \geq 0.5$.
However, since {$B^+ \to \omega K^+$} is a Class V decay and 
$B^+ \to \omega \pi^+$ is a class III decays, they are unstable. 
Thus it is too early to draw definite conclusions.
We expect that
 $B^0 \to \pi^+ \pi^-$,  $\rho^{+} \pi^-$,  
$\rho^+ \rho^-$ and $B^+  \to \pi^+ \pi^0$, $K^{+} \pi^0$ 
will be observed in the next round of experiments
at CLEO and at B factories.

\section{ USEFUL RATIOS }

We start with the  ratios independent of the effective coefficients $a_i$.
Neglecting the small QCD penguin
 contribution and the very small difference in phase space, we get the
relations:
{\begin{eqnarray}
P_2 &\equiv& \frac{{\cal B}(B^0
\to \pi^{-} \pi^+)}{{\cal B} (B^0\to
\rho^{+} \pi^-) }\simeq \left( \frac{f_\pi F_0^{B\to \pi}}
{f_\rho F_1^{B\to \pi}}\right)^2~,\nonumber\\
\label{P3approx}
P_3 &\equiv& \frac{{\cal B}(B^0
\to \pi^{+} \rho^-)}{{\cal B} (B^0\to  
\rho^{+} \pi^-)}\simeq \left(\frac{f_\pi A_0^{B\to \rho}}{
f_\rho F_1^{B\to \pi}}
\right)^2~.
\end{eqnarray}}
 These two ratios $P_2$ and $P_3$ 
can measure $F_0^{B\to \pi}/F_1^{B\to \pi}$ and 
$A_0^{B\to \rho}/F_1^{B\to \pi}$, respectively.
\begin{eqnarray}
P_4 &\equiv&  \frac{{\cal B}({B^+}\to  \pi^+
 \pi^0)}{{\cal B}(B^+\to \rho^+
 \rho^0)} \simeq  \frac{f_\pi^2}{f_\rho^2}\frac{ | F_1^ 
{B\to
\pi}|^2 }{(1+x) |A_1^{B\to \rho}|^2},\nonumber\\
P_{5} &\equiv&  \frac{{\cal B}(B^0
\to \pi^{-} \pi^+)}{{\cal B}(B^0\to
\rho^{-} \rho^+)}\simeq  \frac{f_\pi^2}{f_\rho^2}\frac{ | F_1^ 
{B\to
\pi}|^2 }{(1+x) |A_1^{B\to \rho}|^2},
\end{eqnarray}
where $x=m_\rho /m_B$. The relation $P_4= P_5$ provides a test of 
factorization.
{\begin{eqnarray}
P_1 &\equiv &\frac{{\cal B}(B^0 \to \rho^+ \pi^-) }{{\cal B} (B^0 \to 
\rho^+ \rho^-)}
\simeq \frac{ | F_1^ {B\to \pi}|^2
}{(1+x) |A_1^{B\to \rho}|^2},
\nonumber\\
P_6 &\equiv & \frac{{\cal B}(B^0
\to K^{*+} \pi^-)}{{\cal B}(B^0\to 
K^{*+} \rho^-)}\simeq \frac{ | F_1^ {B\to \pi}|^2
}{(1+x) |A_1^{B\to \rho}|^2}, \nonumber \\
P_7 & \equiv & \frac{{\cal B}({B^+} \to\pi^+
K^{*0})}{{\cal B}(B^+\to \rho^+
 K^{*0} )}\simeq \frac{ | F_1^ {B\to \pi}|^2
}{(1+x) |A_1^{B\to \rho}|^2}.
\end{eqnarray}
They are  essentially determined by the ratios of
 the form factors $F_1^{B\to \pi}$ and $A_1^{B\to \rho}$.
The relation $P_1\simeq P_6 \simeq P_7$ can be a test of factorization.
\begin{table}
\begin{center}\label{pi}
\caption{
 Values of $P_i$'s using the BSW form factors
 and the  lattice-QCD/QCD-sum rule form factors.
The numbers in  brackets are calculated using the approximate
formulae derived in the text.}
\begin{tabular}{ccc}
\hline
Ratio & BSW model & Lattice-QCD \\
\hline
 $P_1$ &  1.19 [1.21] & 1.27 [1.55] \\
 $P_2$ &  0.43 [0.39] & 0.43 [0.39] \\
 $P_3$ &  0.28 [0.28] & 0.27 [0.27] \\
 $P_4$ &  0.49 [0.47] & 0.53 [0.61] \\
 $P_5$ &  0.52 [0.47] & 0.55 [0.61] \\
 $P_6$ &  1.11 [1.21] & 1.19 [1.55] \\
 $P_7$ &  1.11 [1.21] & 1.19 [1.55] \\
 $P_8$ &  1.08 [1.14] & 0.99 [1.18] \\
 $P_9$ &  1.09 [1.14] & 0.99 [1.18]\\
 $P_{10}$ &  1.01 [1.15] & 0.92 [1.19] \\
 $P_{11}$ &  1.01 [1.15] & 0.92 [1.19] \\
\hline
\end{tabular}\end{center}
\end{table}
{\begin{eqnarray}
P_8& \equiv& \frac{{\cal B}({B^+}\to  K^+ \bar K^{*0})}
{{\cal B}(B^+\to K^{*+} \bar K^{*0} )}~,\nonumber\\
P_9 & \equiv &
 \frac{{\cal B}(B^0 \to   K^0
  \bar K^{*0})}{{\cal B}( B^0\to  K^{*0}
 \bar K^{*0} )} ~,\nonumber\\
P_{10} &\equiv& \frac{{\cal B}({B^+}\to  K^+
 \phi)}{{\cal B}(B^+\to K^{*+}
 \phi )} ~,\nonumber\\
P_{11} & \equiv &
 \frac{{\cal B}(B^0 \to  K^0
 \phi)}{{\cal B}( B^0\to  K^{*0}
 \phi )} ~.
\end{eqnarray}}
Ignoring the small phase space difference,
{$$P_8\simeq P_9\simeq P_{10} 
\simeq P_{11} \simeq \frac{ | F_1^ {B\to K}|^2
}{(1+y) |A_1^{B\to K^*}|^2}\label{p4s},$$}
where $y=m_{K^*} /m_B$.
 They are all proportional to the ratios of the form 
factors $F_{1}^{B\to K}$ and $A_1^{B\to K^*}$.
The values of $P_1$-$P_{11}$ calculated in our model
 are displayed in Table~2, together with values calculated using 
the approximate formulae. The two sets of numbers are quite close means that 
the approximations are working quite well.

In the following, we first give ratios to determine the effective coefficients 
 $a_1$ and $a_2$
\begin{eqnarray}
S_1 &\equiv& \frac{{\cal B}(B^0 
\to \pi^+ \pi^-)}{{2}{\cal B}(B^+ \to
\pi^+ \pi^0)}\simeq 
\left(\frac{a_1}{a_1+a_2}\right)^2 ,
\nonumber\\
S_2& \equiv &\frac{{\cal B}(B^0 
\to \rho^+ \rho^-)}{2{\cal B}(B^+ \to 
\rho^+ \rho^0)}\simeq 
\left(\frac{a_1}{a_1+a_2}\right)^2,
\nonumber\\
S_{3}& \equiv& \frac{2{\cal B}(B^+ 
\to \rho^+ \pi^0)}{{\cal B}(B^0 \to 
\rho^+ \pi^-)}\simeq  \left(1+
\frac{1}{x}\frac{a_2}{a_1}\right)^2~,
\nonumber\\
S_{4} &\equiv &\frac{2{\cal B}(B^+ 
\to \pi^+ \rho^0)}{{\cal B}(B^0 \to 
\pi^+ \rho^-)}\simeq  \left(1+
{x}\frac{a_2}{a_1}\right)^2~.
\label{eqs4}
\end{eqnarray}

Similarly for penguin operators,
\begin{eqnarray*}
S_{5} &\equiv &\frac{2{\cal B}(B^+
\to \pi^+ \pi^0)}{{\cal B}(B^+ \to 
\pi^+ K^0)} \simeq \xi^2 \frac{f_\pi^2}{f_K^2} 
\left|\frac{a_1+a_2}{a_4+a_6R_5}\right|^2,
\\
S_{6} &\equiv& \frac{2{\cal B}(B^+ 
\to \rho^+ \rho^{0} )}{{\cal B}(B^+ \to 
\rho^+ K^{*0})} \simeq \xi^2\frac{f_\rho^2 }{f_{K^*}^2} 
\left|\frac{a_1+a_2}{a_4}\right|^2,
\\
S_{7} &\equiv& \frac{{\cal B}(B^0
\to \pi^- \rho^+  )}{{\cal B}( B^+ \to 
\pi^+  K^{*0} )}\simeq \xi^2\frac{f_\rho^2}{f_{K^*}^2}
 \left|\frac{a_1}{a_4} \right|^2 ,  \\
S_{8} &\equiv &\frac{{\cal B}(B^0
\to \rho^- \rho^+  )}{{\cal B}( B^+ \to 
\rho^+  K^{*0} )}\simeq \xi^2\frac{f_\rho^2}{f_{K^*}^2}
 \left|\frac{a_1}{a_4} \right|^2 ,
\\
S_{9} &\equiv& \frac{{\cal B}(B^0
\to \pi^+ \pi^-  )}{{\cal B}( B^+ \to 
\pi^+  K^0 )} \simeq \xi^2\frac{f_\pi^2 }
{f_{K}^2 }  
 \left|\frac{a_1}{a_4+a_6R_5} \right|^2,
\end{eqnarray*}
with $\xi=|V_{ub}V_{ud}^*|/|V_{tb}V_{ts}^*|$.
We can use $S_3,...,S_9$ to determine $a_4$ and $a_6$.
The dominant contribution of the
electroweak penguin amplitudes is proportional to $a_9$. 
It can be determined by the following two ratios.
\begin{eqnarray*}
S_{10} & \equiv & \frac{2{\cal B}(B^0 \to \rho^0 K^0)}
{{\cal B}(B^+ \to \pi^+ K^{*0} )} \simeq \frac{9}{4} \left|
\frac{f_\rho F_1^{B \to 
K}}{f_{K^*} F_1^{B \to \pi}}\right|^2 
\left|\frac{a_9}{a_4}\right|^2,
\nonumber\\
S_{11} & \equiv & \frac{2{\cal B}(B^0 \to \rho^0 K^0)}
{{\cal B}(B^0 \to \rho^-\pi^+ )} \simeq \frac{9}{4\xi^2}
 \left|\frac{f_\rho 
F_1^{B \to
K}}{f_\pi A_0^{B \to \rho}}\right|^2 
\left|\frac{a_9}{a_1}\right|^2.
\end{eqnarray*}
It is difficult to quantitatively determine 
the other penguin coefficients which are smaller.

 The ratio discussed by
Fleischer and Mannel recently to constrain the CKM parameter 
$\gamma$ \cite{fm}, is defined as
\begin{eqnarray}
\label{S12}
S_{12} &\equiv& \frac{{\cal B}(B^0 
\to  K^+ \pi^- )}{{\cal B}(B^+ \to
K^0 \pi^+ )} \\ \nonumber
&\simeq& 1-2z_{12}\cos \delta_{12} \cos \gamma +z_{12}^2,
\end{eqnarray}
where $\delta_{12}$ is the strong phase, and 
$z_{12}$ is defined in ref.~\cite{akl1}.
 The ratio
$S_{12}$  has the experimental value:
$S_{12}= 1.0 \pm 0.4  $.
It is shown in Fig.~\ref{ss1}.
We see that the measurement of $S_{12}$ can determine the CKM
parameter $\cos \gamma$.
Analogous to eqn.~(\ref{S12}), we  define 
\begin{eqnarray}
S_{13} &\equiv &\frac{{\cal B}(B^0 
\to \pi^- K^{*+} )}{{\cal B}(B^+ \to 
\pi^+ K^{*0})}\nonumber\\
S_{15} &\equiv &\frac{{\cal B}(B^0 
\to \rho^- K^{*+} )}{{\cal B}(B^+ \to 
\rho^+ K^{*0})}\\ \nonumber
S_{13}\simeq  S_{15}  
&\simeq & 1-2z_{13} \cos \delta_{13} \cos \gamma +z_{13}^2.
\end{eqnarray}
They can also be used to determine the $\cos \gamma$.
\begin{figure}
    \epsfig{file=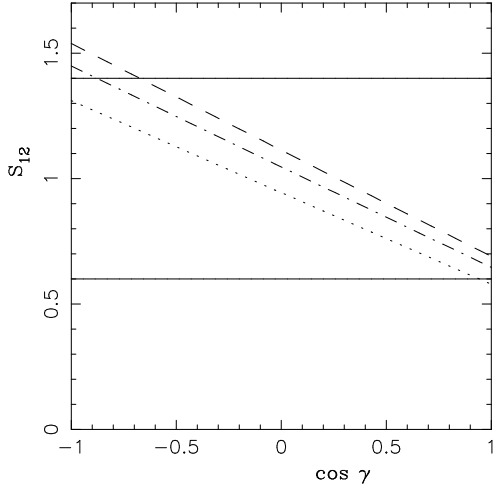,bbllx=5.4cm,bblly=8.4cm,bburx=21cm,bbury=20cm,%
width=8.3cm,angle=0}
    \caption{
$S_{12}$ as a function of $\cos \gamma$.
The dotted, dash-dotted and dashed curves correspond to $N_c=\infty$,   
$N_c=3$ and $N_c=2$, respectively. The horizontal lines are the CLEO
$(\pm 1\sigma)$ measurements of $S_{12}$.
}
\label{ss1}
\end{figure}

\section{ SUMMARY}

 The recently measured charmless nonleptonic B decay modes 
 can be explained in the factorization framework. 
They show some preference for $\xi \leq 
0.2$.   A good
fraction of the seventy six decay modes  will be measured in the future
providing a detailed test of the factorization approach. 

  We have put forward numerous proposals for
 ratios of branching ratios
to determine the effective coefficients {$a_1,a_2,a_4,
a_6$ and $a_9$}.
 We  proposed also a number of ratios $P_i$, 
 which will help in determining the form factors for the 
various decays considered here. 
 The 
two-body nonleptonic decays  provide potentially
non-trivial constraints on the CKM parameters.

 Finally,   it is
instructive to study direct and indirect CP violation in all the two-body
nonleptonic $B$ decays considered here \cite{akl2}.

\end{document}